%% file: fmcw_Jcas.tex
\documentclass[conference,10pt,letterpaper]{IEEEtran}
\IEEEoverridecommandlockouts
\usepackage{cite}
\usepackage{amsmath,amssymb,amsfonts, mathdots, bm}
\usepackage{algorithmic}
\usepackage{graphicx}
\usepackage{textcomp}
\usepackage{xcolor}
\usepackage{color}
\usepackage{siunitx}
\usepackage{todonotes}
\usepackage{tikz-3dplot}
 \usepackage{mleftright}
\usepackage{pgf, tikz, pgfplots}
\usetikzlibrary{external}
\pgfplotsset{compat=1.18}
\usepackage{circuitikz}
\usetikzlibrary{shapes,arrows, calc, fit}
\usetikzlibrary{positioning}
\usetikzlibrary{fit}
\def\BibTeX{{\rm B\kern-.05em{\sc i\kern-.025em b}\kern-.08em
    T\kern-.1667em\lower.7ex\hbox{E}\kern-.125emX}}
\usepackage[numbers]{natbib}

\usepackage{adjustbox}
\usepackage{svg}
\usepackage{lipsum}
\usepackage{natbib}
\usepackage[nolist]{acronym}
\usepackage{pgfplots}
\usepackage{pgfplotstable}
\usepackage{cite}
\usepackage{dblfloatfix}

\input{beamercolorthemekit.tex}

\newcommand{\RV}[1]{\mathsf{#1}}

\def\j{{\mathrm{j}}}
\def\e{{\mathrm{e}}}

\newcommand\blfootnote[1]{%
  \begingroup
  \renewcommand\thefootnote{}\footnote{#1}%
  \addtocounter{footnote}{-1}%
  \endgroup
}

\begin{document}
\input{acronyms.tex}
\title{On the Sensing Performance of FMCW-based Integrated Sensing and Communications with Arbitrary Constellations
}

\author{Daniel Gil Gaviria, Benedikt Geiger, Charlotte Muth, Laurent Schmalen\\
\IEEEauthorblockA{Communications Engineering Lab (CEL), Karlsruhe Institute of Technology (KIT) \\ Hertzstr. 16, 76187 Karlsruhe, Germany, Email: \texttt{daniel.gil@kit.edu}}
}

\maketitle

\begin{abstract}
    Integrated sensing and communications (ISAC) is expected to play a major role in numerous future applications, e.g., smart cities. Leveraging native radar signals like the frequency modulated continuous wave (FMCW) waveform additionally for data transmission offers a highly efficient use of valuable physical radio frequency (RF) resources allocated for automotive radar applications. In this paper, we propose the adoption of higher-order modulation formats for data modulation onto an FMCW waveform and provide a comprehensive overview of the entire signal processing chain. We evaluate the impact of each component on the overall sensing performance. While alignment algorithms are essential for removing the information signal at the sensing receiver, they also introduce significant dispersion to the received signal. We analyze this effect in detail. Notably, we demonstrate that the impact of non-constant amplitude modulation on sensing performance is statistically negligible when the complete signal processing chain is considered. This finding highlights the potential for achieving high data rates in FMCW-ISAC systems without compromising the sensing capabilities.
\end{abstract}

\section{Introduction}
\input{1introduction}

\section{System Model}
\input{2System_Model}

\section{Performance Analysis}\label{sec:Performance}

\input{3Performance_Analysis.tex}

\section{Numerical Results}

\input{4results.tex}

\section{Conclusion}
\input{5conclusions.tex}

\bibliographystyle{IEEEtran}
\bibliography{references_fmcw.bib}

\end{document}

%% file: beamercolorthemekit.tex
\definecolor{kit-gray70}{cmyk}{.6,.6,.6,0}

\definecolor{kit-blue100}{cmyk}{.8,.5.,0,0}
\definecolor{kit-blue70}{cmyk}{.56,.35,0,0}
\definecolor{kit-blue50}{cmyk}{.4,.25,0,0}
\definecolor{kit-blue30}{cmyk}{.24,.15,0,0}
\definecolor{kit-blue15}{cmyk}{.12,.075,0,0}

\definecolor{kit-green100}{cmyk}{1,0,.6,0}
\definecolor{kit-green70}{cmyk}{.7,0,.42,0}
\definecolor{kit-green50}{cmyk}{.5,0,.3,0}
\definecolor{kit-green30}{cmyk}{.3,0,.18,0}
\definecolor{kit-green15}{cmyk}{.15,0,.09,0}

\definecolor{KITgreen}{rgb}{0,.59,.51}

\definecolor{KITpalegreen}{RGB}{130,190,60}
\colorlet{kit-maigreen100}{KITpalegreen}
\colorlet{kit-maigreen70}{KITpalegreen!70}
\colorlet{kit-maigreen50}{KITpalegreen!50}
\colorlet{kit-maigreen30}{KITpalegreen!30}
\colorlet{kit-maigreen15}{KITpalegreen!15}

\definecolor{KITblue}{rgb}{.27,.39,.66}

\definecolor{KITyellow}{rgb}{.98,.89,0}
\definecolor{kit-yellow100}{cmyk}{0,.05,1,0}
\definecolor{kit-yellow70}{cmyk}{0,.035,.7,0}
\definecolor{kit-yellow50}{cmyk}{0,.025,.5,0}
\definecolor{kit-yellow30}{cmyk}{0,.015,.3,0}
\definecolor{kit-yellow15}{cmyk}{0,.0075,.15,0}

\definecolor{KITorange}{rgb}{.87,.60,.10}
\definecolor{kit-orange100}{cmyk}{0,.45,1,0}
\definecolor{kit-orange70}{cmyk}{0,.315,.7,0}
\definecolor{kit-orange50}{cmyk}{0,.225,.5,0}
\definecolor{kit-orange30}{cmyk}{0,.135,.3,0}
\definecolor{kit-orange15}{cmyk}{0,.0675,.15,0}

\definecolor{KITred}{rgb}{.63,.13,.13}
\definecolor{kit-red100}{cmyk}{.25,1,1,0}
\definecolor{kit-red70}{cmyk}{.175,.7,.7,0}
\definecolor{kit-red50}{cmyk}{.125,.5,.5,0}
\definecolor{kit-red30}{cmyk}{.075,.3,.3,0}
\definecolor{kit-red15}{cmyk}{.0375,.15,.15,0}

\definecolor{KITpurple}{RGB}{160,0,120}
\colorlet{kit-purple100}{KITpurple}
\colorlet{kit-purple70}{KITpurple!70}
\colorlet{kit-purple50}{KITpurple!50}
\colorlet{kit-purple30}{KITpurple!30}
\colorlet{kit-purple15}{KITpurple!15}

\definecolor{KITcyanblue}{RGB}{80,170,230}
\colorlet{kit-cyanblue100}{KITcyanblue}
\colorlet{kit-cyanblue70}{KITcyanblue!70}
\colorlet{kit-cyanblue50}{KITcyanblue!50}
\colorlet{kit-cyanblue30}{KITcyanblue!30}
\colorlet{kit-cyanblue15}{KITcyanblue!15}

%% file: acronyms.tex
\begin{acronym}[]
    \acro{6G}{sixth generation}
    \acro{AE}{autoencoder}
    \acro{AF}{ambiguity function}
    \acro{AIR}{achievable information rate}
    \acro{AWGN}{additive white Gaussian noise}
    \acro{CA}{cell-averaging}
    \acro{CFAR}{constant false alarm rate}
    \acro{CP}{cyclic prefix}
    \acro{FEC}{forward error correction}
    \acro{FFT}{fast Fourier transform}
    \acro{GMI}{generalized mutual information}
    \acro{IFFT}{inverse fast Fourier transform}
    \acro{ISAC}{integrated sensing and communications}
    \acro{LLR}{log-likelihood ratio}
    \acro{MF}{matched filter}
    \acro{OFDM}{orthogonal frequency division multiplexing}
    \acro{PSK}{phase shift keying}
    \acro{RCS}{radar cross section}
    \acro{QAM}{quadrature amplitude modulation}
    \acro{RV}{random variable}
    \acro{SC}[S\&C]{sensing \& communications}
    \acro{SD}{soft decision}
    \acro{SINR}{signal-to-interference-and-noise ratio}
    \acro{SNR}{signal-to-noise ratio}
    \acro{TOI}{target of interest}
    \acro{FMCW} {Frequency modulated continuous wave}
    \acro{PC-FMCW} {phase coded FMCW}
    \acro{ADC} {analog-to-digital converter}
    \acro{RVPC}{residual video phase compensation}
    \acro{IF}{intermediate frequency}
    \acro{ISNR}{image signal-to-noise-ratio}
    \acro{RDM}{range Doppler matrix}
\end{acronym}

%% file: 1introduction.tex
Future smart cities are anticipated to integrate autonomous driving, intelligent public traffic management, and high vehicular connectivity, among other features. These innovations will depend heavily on precise environmental awareness and efficient communication among a large network of interconnected nodes. \Ac{ISAC} offers a game-changing solution to realize these visions. By leveraging hardware and software resources for both sensing and communications, \ac{ISAC} maximizes the utilization of costly physical resources, such as transmission power and RF-frequency bands, significantly enhancing the overall system efficiency.\blfootnote{This work received funding from the German Federal Ministry for Economic Affairs and Climate Action (BMWK) under grant agreement 20Q1964B and from the the German Federal Ministry for Education and Research (BMBF) under grant agreements 16KISK010 (Open6GHub) and 16KISK123 (KOMSENS-6G).}

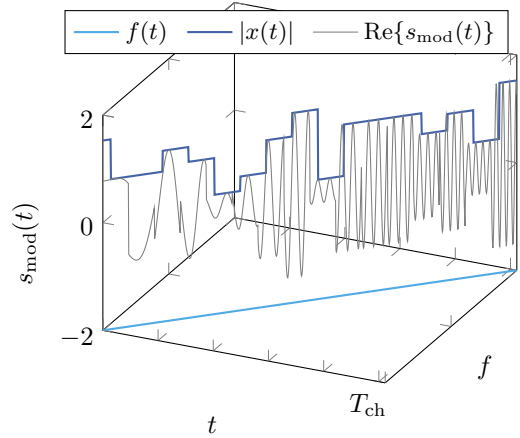
\begin{figure}
    \input{figures/fmcw_skizze} 
    \caption{Sketch of the absolute value and instantaneous frequency of an {FMCW-ISAC} transmit signal modulated with 64-QAM symbols.}
    \label{fig:fmcw_sketch}   
\end{figure}
While significant progress is taking place in integrating advanced sensing capabilities into the latest cellular communication networks \cite{liu22, shatov}, another promising avenue is the use of automotive radar waveforms to transmit communication signals. Radar waveforms not only offer high ranging accuracy due to their wide bandwidths but also provide additional opportunities for data transmission. The short range radar (SRR) band between \SI{77}{\giga\hertz} and \SI{81}{\giga\hertz} is of particular interest. The rapid advances in semiconductor processing in the millimeter wave range allow for efficient and reliable signal generation and processing through compact RF-front ends at affordable production costs for industrial mass production \cite{walschmidt_review_2021}.

\renewcommand{\thefigure}{3}
\begin{figure*}[b]
    \input{figures/fmcw_RVPC_phase_coded.tex} 
    \caption{Block diagram of a complete FMCW-ISAC system. Components of a traditional FMCW radar shown in white, additional blocks for communication capabilities are shown in blue.} 
    \label{fig:fmcw_isac_block}
\end{figure*}
\Ac{FMCW} is one of the most widely spread state-of-the-art waveforms for broadband radar applications. Its primary advantage lies in its reliance on mixed-signal processing, which enables large bandwidth coverage in the RF domain while maintaining simple and low-cost hardware requirements. Specifically, the processing of beat frequencies requires relatively low sampling rates, as these frequencies are significantly smaller than the sweep bandwidth of the transmitted signal. This reduces the cost of the \ac{ADC} and, consequently, the overall system costs. An exemplary FMCW-ISAC waveform is sketched in Fig. \ref{fig:fmcw_sketch}.

\renewcommand{\thefigure}{2}
\begin{figure}
    \input{figures/ber_M16.tex}
    \caption{Exemplary bit error rates (BER) for 16-PSK and 16-QAM in an AWGN-channel.}
    \label{fig:ber}
\end{figure}

It is well known that every \ac{ISAC} system has to fundamentally  deal with a trade-off between sensing and communication performance \cite{liu_2023_fundamental}. In general, signals optimized for radar such as the traditional FMCW waveform have a constant envelope over time and frequency. In contrast, communication systems seek to maximize the mutual information between the transmit and receive signals, typically utilizing both phase and amplitude as degrees of freedom. 
For this reason, \ac{PC-FMCW}  has emerged as a natural and widely adopted solution for FMCW-ISAC systems \cite{Lampel_2022}. On the other hand, the use of higher order modulations with non-constant magnitude might enable a more efficient use of the communication capabilities. For instance, Fig. \ref{fig:ber} shows how the use of magnitude as a degree of freedom for communications (16-QAM) through an AWGN channel leads to several dB gain in terms of BER performance. The impact of this kind of modulation on the sensing performance of FMCW-ISAC systems has not been studied in detail in the literature. 

FMCW-ISAC requires additional signal processing efforts as compared to classical automotive radar, especially at the sensing receiver. More specifically, state of the art \ac{PC-FMCW} radar rely on the \ac{RVPC} algorithm \cite{Meta_2007} to align and compensate the phase modulation from the received echoes. This algorithm has also been studied for interference mitigation in MIMO scenarios in \cite{Uysal_2020,Petrov_2023}. The use of \ac{RVPC} for \ac{ISAC} applications was considered in \cite{Lampel_2022}.  Although these publications qualitatively observe a degradation of the sensing performance due to the dispersion caused by the alignment algorithm,  a quantitative evaluation of both sensing and communication performance metrics still needs to be derived. Additionally, none of the prior works has evaluated the effect of higher order modulation formats with non-constant modulus (e.g. QAM) on the performance of the overall system.

In this paper, we provide following contributions:
    We give a comprehensive overview of the signal processing chain for FMCW-ISAC with modulations of arbitrary order.
    We present a quantitative study on the effect of the RVPC algorithm on the evaluated signal and its impact the sensing performance.
    Finally, we compare the performance of FMCW-ICAS-systems using purely phase coded modulation formats and higher order QAM-modulations. In fact we show that the impact of higher order modulations on the sensing is statistically negligible if the complete processing chain is considered.

%% file: figures/fmcw_skizze.tex
\centering
\begin{tikzpicture}[]
    \begin{axis}[
        width=.8\columnwidth, 
        height=.75\columnwidth, 
        xlabel=$t$,
        ylabel=$f$,
        zlabel=$s_\mathrm{mod}(t)$, 
        zmin=-2,
        zmax=2,
        xtick={0, 200, ..., 1000},
        xticklabels={, , , , ,$T_\mathrm{ch}$},
        ytick={0, 0.5,1}, 
        yticklabels={, , $B$},
        legend style={
        anchor=north east,
        legend columns=-1,
        /tikz/every even column/.append style={column sep=0.5em},
        font=\small
    }
    ]
        \addplot3 [thick, KITcyanblue] table [x=t, y=f, z expr=-2, col sep=comma, mark=none,] {figures/tx_3dPlot_64.csv};     
        \addlegendentry{$f(t)$}   
        \addplot3 [thick, KITblue]table [x=t, y=f, z=abs, col sep=comma, mark=none, ] {figures/tx_3dPlot_64.csv};
        \addlegendentry{$|x(t)|$}   
        \addplot3 [thin, gray]table [x=t, y=f, z=real, col sep=comma, mark=none, ] {figures/tx_3dPlot_64.csv};
        \addlegendentry{$\mathrm{Re}\mleft\{s_\mathrm{mod}(t)\mright\}$}   
    \end{axis}
    \end{tikzpicture} 

%% file: figures/fmcw_RVPC_phase_coded.tex
\centering

\tikzset{
    ultra thin/.style= {line width=0.1pt},
    very thin/.style=  {line width=0.2pt},
    thin/.style=       {line width=0.4pt},%
    semithick/.style=  {line width=0.6pt},
    thick/.style=      {line width=0.8pt},
    very thick/.style= {line width=1.2pt},
    ultra thick/.style={line width=1.6pt}
}

    \tikzstyle{bl} = [draw, line width=1pt, rectangle, inner sep=.5em, rounded corners, fill=kit-cyanblue30, font=\footnotesize]
    \tikzstyle{bl_white} = [draw, line width=1pt, rectangle, inner sep=.5em, rounded corners, font=\footnotesize]
    \tikzstyle{bl2} = [draw, very thick, circle, inner sep=.8em]
     \tikzstyle{input} = [coordinate]
     \tikzstyle{output} = [coordinate]
     \tikzstyle{every node}=[line width=1pt, font=\footnotesize]
     \tikzstyle{dot}=[circle,fill,draw,inner sep=0pt,minimum size=1pt]
    \tikzstyle{sum}=[draw, circle, minimum size=10pt]
    \tikzset{>=latex}
\begin{circuitikz}[auto, line width=1pt, x=0.2cm,y=0.4cm]
    
    \node[oscillator, name=chirp, align=center, scale=0.4,line width=0.6pt]{};
    \node[mixer,  fill=kit-cyanblue30, line width=0.6pt, scale=0.45,name=mix1, align=center, above right= 1.5 and 2 of chirp]{};
    \node[left=0.5 of chirp, align=center, name=uj]{$s(t)$};

    \node[bareantenna,line width=0.6pt, scale=0.4, above = 1.5 of mix1, name=ant_out] {};
    \node[right=0.3 of ant_out]{Tx};
    \node[bareantenna, line width=0.6pt, scale=0.4, right =6 of ant_out, name= ant_in]{};
    \node[right=0.5 of ant_in]{Rx};
    \node[mixer, line width=0.6pt, scale=0.45, below=6 of ant_out,name=mix]{};
   \node[bl_white, left=6 of mix, align=center, name=dac](dac){ADC};
    \node[bl, left=6 of dac, align=center, name=rvpc](rvpc){Alignment};
    \node[bl, left= 5 of rvpc,name=Compensation, align=center]{Compensation};
    \node[bl_white, left=8 of Compensation, name=RDM, align=center]{RDM \\ calculation};
    
    \node [bl, name=adc, left= 6 of mix1] {DAC} ; 
    \node[bl, name=ps, left= 30 of adc, align=center,name=ps]{Pulse \\ shaping};
    \node[left=4 of ps, align=center, name=xj]{$\RV{c}_m$};
    \node[bl, name=tx_rvpc, above=1 of Compensation,  align=center]{Alignment};
 
    \draw [arrows = {-Latex[width = 1.5mm]}] (chirp)-|node[name=x, below] {} (mix1);
    \draw[->] (chirp)-|(mix); 
    \draw[arrows = {-Latex[width = 1.5mm]}] (mix1) to [short, -*] node[name=x, below] {} (chirp-|mix1)--  (mix1);
    \draw[-] (mix1) -- node[right]{$s_\mathrm{mod}(t)$}(ant_out);
    \draw[->] (ant_in)|-node[name=x, below] {$r(t)$}(mix);
    \draw[arrows = {-Latex[width = 1.5mm]}] (mix)--node[below]{$r_\mathrm{IF}(t)$}(dac);
    \draw[arrows = {-Latex[width = 1.5mm]}] (dac)--node[below]{$r_\mathrm{IF}[n]$} (rvpc);
    \draw[->] (rvpc) -- node[below]{$r_\mathrm{al}[n]$}(Compensation);

    \draw[->] (ps) to[short, -*](xj-|Compensation) to(tx_rvpc);
    \draw[->](tx_rvpc)-- node[right]{$\tilde{x}[n]$}(Compensation);
    \draw[->](xj)--(ps);
    \draw[->](ps)--(adc);
    \draw[->] (adc)-- node[above]{$x(t)$}(mix1);
    \draw[->] (Compensation) --node[below]{$r_\mathrm{comp}[n]$}(RDM);
    
    \node[draw,dashed,KITgreen, fit= (rvpc) (tx_rvpc) (Compensation), inner sep=0.4em](rvpc_box) {};
    \draw[] (rvpc_box.north east) node[below left, KITgreen]{RVPC};

    \end{circuitikz}

%% file: figures/ber_M16.tex
\centering
\begin{tikzpicture}
    \begin{axis}[width=0.7\columnwidth,
      ylabel={$\mathrm{BER}$},
      xlabel={$E_\mathrm{b}/N_0$}, 
      ymode=log,
      legend cell align={left},
      legend style={
         at={(0.05,0.05)},
        anchor=south west,
        legend columns=1,
        mark size=.6pt,
        font=\small,
    },
      xmin=-10,
      xmax=20,
      ymin=1e-5,
      grid=major
      ]

      \addplot [KITblue, thick]
      table[x=snr,y=16psk, col sep=comma] {figures/ber_M16.csv};
      \addlegendentry{16-PSK};
      \addplot [KITgreen, thick, dashed]
      table[x=snr, y=16qam, col sep=comma] {figures/ber_M16.csv};
      \addlegendentry{16-QAM};
    \end{axis}

  \end{tikzpicture}

%% file: 2System_Model.tex
This paper considers a monostatic FMCW-ISAC system as illustrated in Fig. \ref{fig:fmcw_isac_block}. In such a system, the transmitted signal is fully known to the sensing receiver, eliminating the need for additional synchronization for sensing purposes. This section explains how communication capabilities are incorporated into a classical \ac{FMCW} radar system and describes the additional sensing signal processing steps required when random communication symbols are modulated onto the transmitted signal.

\subsection{FMCW-ISAC Transmitter}
A classical \ac{FMCW} radar transmits chirps
\begin{align}
    s(t)&=\e^{\j 2\pi \left(f_\mathrm{c}t+\frac{1}{2} \alpha t^2\right)}\text{,}& 0 \leqslant &t<T_\mathrm{ch} \text{,}
\end{align}
with a linear frequency slope $\alpha = B_\mathrm{ch}/T_\mathrm{ch}$, where $f_\mathrm{c}$ denotes the carrier frequency, $B_\mathrm{ch}$ represents the chirp bandwidth, and $T_\mathrm{ch}$ is the chirp duration.

To integrate communication capabilities, a communication signal is modulated onto $s(t)$. For each chirp, the information source emits $M$ random complex modulation symbols $\RV{c}_m$ drawn from a unit power modulation alphabet, i.e., ${\RV{c}_m \in \mathcal{C} \subset \mathbb{C}}$, where $\mathbb{E}\{| \RV{c}| ^2 \}=1$. The baseband communication signal
\begin{align}
    x(t)=\sum_{m=0}^{M-1}\RV{c}_m g_\mathrm{ps}\left(t-m\cdot\frac{T_\text{ch}}{M}\right)
\end{align}
consists of the $M$ modulation symbols of duration $\frac{T_\mathrm{ch}}{M}$, shaped by a pulse shaping filter with impulse response $g_\mathrm{ps}(t)$, such as a root-raised-cosine filter. 

The communication signal $x(t)$ is then modulated onto the raw chirp $s(t)$ yielding the modulated \ac{FMCW} signal
\begin{equation}
        s_\mathrm{mod}(t)= x(t) \cdot s(t) = x(t) \cdot \e^{\j 2\pi \left(f_\mathrm{c}t+\frac{1}{2} \alpha t^2\right)} \text{.}
\end{equation}

Depending on the modulation alphabet $\mathcal{C}$, three cases can be distinguished:
\begin{itemize}
    \item No modulation, i.e., $\RV{c}_m=1, \forall m$, results in a traditional \ac{FMCW} radar system without communication capabilities.
    \item A \ac{PSK} modulation, i.e., $c_i = \e^{\j \varphi _i}, \forall i$, such as BPSK or QPSK, results in a \ac{PC-FMCW} system studied by \cite{Petrov_2023} and \cite{Lampel_2022}.
    \item A higher order QAM modulation (e.g. 16-QAM or 64-QAM) results in a varying amplitude of $s_\mathrm{mod}(t)$. In order to enforce unity average power over the duration of each chirp, algorithms like constant composition distribution matching can be applied \cite{schulte2016}.
\end{itemize}
This paper focuses on investigating the use of QAM within the \ac{FMCW}-ISAC framework.

\subsection{Sensing Channel}
The sensing channel $h(t)$ can be modelled as a superposition of $L$ independent reflections 
    \begin{align*}
        h(t)&=\sum_{\ell=0}^{L-1}a_\ell\delta(t-\tau_\ell)\cdot\e^{\j2\pi f_{\mathrm{D},\ell}t}\text{,}
    \end{align*}
    with an amplitude $a_\ell$, a time delay $\tau_\ell$ and doppler shift $f_{\mathrm{D},\ell}$. The received signal is then given by
    \begin{align*}
        r(t) &= \sum_{\ell=0}^{L-1} a_\ell s_\mathrm{mod}(t-\tau_\ell) \, \e^{\,\j 2\pi f_{\mathrm{D},\ell} t } + w(t)\\
            &=\sum_{\ell=0}^{L-1} a_\ell x(t-\tau_\ell)\e^{\j 2\pi \left(f_c (t-\tau_\ell)+\frac{1}{2} \alpha (t-\tau_\ell)^2 + f_{\mathrm{D}, \ell} t\right)}+w(t)\text{.}
     \end{align*}
     The \ac{AWGN} is represented by $w(t)\sim\mathcal{CN} \mleft(0, \sigma^2\mright)$. The signal is then downconverted using the complex conjugate of the chirp signal $s(t)$ used for upconversion. The resulting signal $r_\mathrm{IF}(t)=r(t)s^*(t)$ is referred to as \ac{IF} signal
    \begin{align}\label{eq:IF_signal_t}
        r_\mathrm{IF}(t) 
                        &=\sum_{\ell=0}^{L-1} a_\ell \underbrace{x(t-\tau_\ell)}_{\substack{\text{delayed}\\\text{sequence}}} \cdot \underbrace{\e^{\j 2\pi \varphi_\ell}}_{\substack{\text{phase}\\\text{shift}}} \cdot \underbrace{\e^{-\j 2\pi\alpha \tau_\ell t}}_{\substack{\text{beat}\\ \text{signal}}}
                        \underbrace{\e^{\j2\pi f_{\mathrm{D},\ell}}}_{\substack{\text{Doppler}\\ \text{Shift}}}+\check{w} (t)\text{,}
    \end{align}
     where $\check{w}(t)$ represents the \ac{AWGN} with unchanged statistical characteristics. Note that each target manifests as a delayed version of the information signal $x(t-\tau_\ell)$, a constant phase shift ${\varphi_\ell=f_\text{c} \tau_\ell-\frac{1}{2} \alpha \tau_\ell^2}$, a complex oscillation with beat frequency
    \begin{equation}\label{eq:f_b-tau}
        f_{\text{beat},\ell}=-2\pi\alpha\tau_\ell\text{,}
    \end{equation}
    and an additional Doppler shift $f_{\mathrm{D},\ell}$. 
    \renewcommand{\thefigure}{4}
    \begin{figure}
        \input{figures/spectrogram.tex}
        \caption{Spectrogram for a system with $T_\mathrm{ch}=1024T_\mathrm{s}$ with one target at $\tau=256 T_\mathrm{s}$ and $SNR=\SI{30}{\decibel}$ before signal processing. } 
        \label{fig:spectrogram}
    \end{figure}
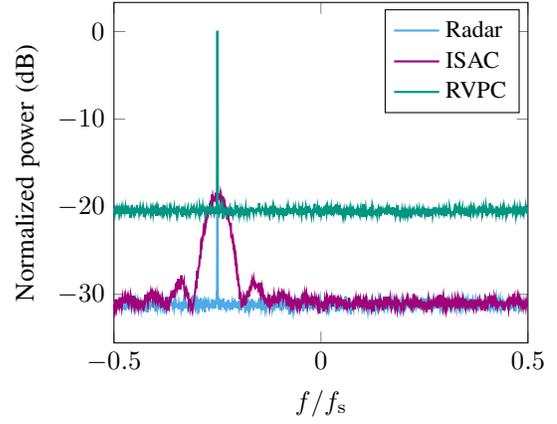

Next,  the IF signal $r_\mathrm{IF}(t)$ is sampled and all subsequent steps are carried out in the digital domain. 
The discrete-time representation of (\ref{eq:IF_signal_t}) is given by
\begin{equation}\label{eq:IF_signal_n}
    r_\mathrm{IF}[n] %
                    =\sum_{\ell=0}^{L-1} a_\ell x\mleft[n-\frac{\tau_\ell}{T_\mathrm{s}}\mright] 
                    \e^{\j 2\pi \varphi_\ell} 
                    \e^{\j 2\pi T_\mathrm{s}  (-f_{\mathrm{beat, }\ell}+f_\mathrm{D, }\ell) n}+\check{w}[n]\text{,}
\end{equation}
where $T_\mathrm{s}$ is the sampling period and $n$ is the discrete time variable. This results in a total of $N=\frac{T_\mathrm{ch}}{T_\mathrm{s}}$ samples per chirp. 

 In a traditional FMCW radar scenario, where $x[n]=1, \forall n$, only the complex oscillations with frequencies $-f_{\text{beat},\ell}$ and $f_{\mathrm{D},\ell}$ remain. These two frequencies are proportional to the range and velocity of the targets, respectively.  To estimate these parameters separately, a \ac{RDM} is computed using an FFT across the samples of a single chirp, followed by an \ac{IFFT} over several consecutive chirp repetitions as detailed in \cite{winkler2007}.

 Broadband FMCW systems are typically designed such that $f_{\mathrm{beat},\ell} \gg f_{\mathrm{D},\ell}$, rendering the impact of $f_{\mathrm{D},\ell}$ on the overall system performance negligible. For this reason, we will neglect $f_{\mathrm{D}, \ell}$ and focus solely on estimating $f_{\mathrm{beat},\ell}$. This step is done through a single FFT over the duration of a single chirp and yields a periodogram given by
 \begin{equation}\label{eq:RDM}
    \mathrm{Per}[\hat{f}_\mathrm{beat}] =  \mathrm{FFT} \left\{ r_\mathrm{IF}[n] \right\} .
 \end{equation}
 Additionally, we assume that the power of each target is perfectly concentrated within a single bin of the periodogram. While these assumption may not always hold in real physical systems, the precise position of the peak can still be accurately estimated through interpolation methods \cite{richards2005}. An exemplary periodogram with a single target is shown in Fig. \ref{fig:spectrogram}. The blue line is obtained for using the native FMCW signal with no communication capabilities.

\subsection{RVPC-Alignment algorithm}
In our \ac{ISAC} scenario, the multiplication by $x(t-\tau_\ell)$ spreads the spectrum of $r_\mathrm{IF}[n]$, making it difficult to directly estimate $\hat{f}_{\text{beat},\ell}$ or $\hat{\tau}_\ell$. This phenomenon is illustrated in \nolinebreak{Fig.~\ref{fig:spectrogram}}. The purple curve shows the spectrum of  $r_\mathrm{IF}[n]$, which is spread over the frequency domain and shaped as the spectrum of $x[n]$. At the same time, an estimate of all $\tau_\ell$ is required to remove the information signal $x[n - \frac{\tau_\ell}{T\mathrm{s}}]$ from $r_\mathrm{IF}[n]$, resulting in a chicken-and-egg problem. To resolve this issue, the \ac{RVPC} algorithm employs a frequency-domain group delay filter to align all potential frequency components of $r_\mathrm{IF}[n]$ without prior knowledge of their delays, leveraging the entanglement between $f_{\mathrm{beat},\ell}$ and $\tau_\ell$.
\renewcommand{\thefigure}{5}
\begin{figure}
    \input{dispersion_sketch}
    \caption{Sketch of the alignment filter $G_\mathrm{al}(f)$ and the dispersion effect on the aligned signals.}
    \label{fig:dispersion_sketch}
\end{figure}
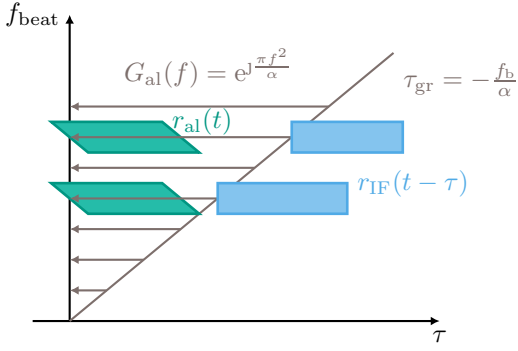

Observe that $f_{\mathrm{beat}, l}=\tau_\ell\alpha$ holds for all summands $\ell$ in (\ref{eq:IF_signal_t}). This relation is shown in Fig. \ref{fig:dispersion_sketch}, where an ISAC signal $r_\mathrm{IF}(t)$ with $L=2$ target is shown in the form of blue rectangles.  To align these components, an allpass alignment filter $G_\mathrm{al}(f) =\e^{\j \varphi_\mathrm{al}(f)}$ must introduce a group delay that compensates for the position of each frequency component. Therefore, its phase response $\varphi_\mathrm{al}(f)$  is designed to match the group delay according to
\begin{align}
    \tau_\mathrm{gr}&=-\frac{\mathrm{d}\varphi_\mathrm{al}(f)}{2\pi \mathrm{d}f}\stackrel{!}{=}-\frac{f}{\alpha}\text{,}\\
    \varphi_\mathrm{al}(f)&=\pi\frac{f^2}{\alpha}\text{.}
\end{align} 
 The discrete time representation of $G_\mathrm{al}[k]$ as an acausal discrete filter is given by
\begin{align} 
    G_\mathrm{al}[k] = \e^{\j \frac{k^2}{\alpha}} \text{,} 
\end{align} 
where $k$ is the discrete frequency index. To apply $G_\mathrm{al}[k]$ in the discrete frequency domain, the following steps are performed: an FFT of $r_\mathrm{IF}[n]$ is computed, followed by an elementwise multiplication with $G_\mathrm{al}[k]$, and subsequently an IFFT is applied to get $r_\mathrm{al}[n]$ in time domain as a result:
\begin{equation}\label{eq:G_implementation}
    r_\mathrm{al}[n]=\mathrm{IFFT}\mleft\{\mathrm{FFT}\mleft\{ r_\mathrm{IF}[n] \mright\} \cdot G_\mathrm{al}[k]\mright\}{\text{.}}
 \end{equation}

 The aligned signal $r_\mathrm{al}(t)$ at the output of the filter is shown in green in Fig. \ref{fig:dispersion_sketch}. Note that $G_\mathrm{al}[k]$ introduces dispersion, as higher frequency components experience a larger group delay compared to low frequency components, leading to the illustrated parallelogram shape. This effect becomes particularly relevant for signals $x[n]$ with high symbol rates and correspondingly large bandwidth. As a result, the aligned signal $r_\mathrm{al}[n]$ can be written as
   \begin{align}
    r_\mathrm{al}[n]=
    \tilde{x}\left[n\right] \sum_{l=0}^{L-1} a_\ell \cdot
    \e^{\j 2\pi \varphi_\ell}\cdot 
    \e^{\j \frac{2\pi}{f_\mathrm{S}}  \alpha \tau_\ell n}+\breve{w}[n]\text{,}
\end{align}
where $\tilde{x}[n]$ and $\breve{w}[n]$ are the dispersed versions of $x[n]$ and $\check{w}[n]$ respectively. Since $G_\mathrm{al}[k]$ is an allpass filter, the stochastic characteristics of the noise remain unchanged.

Since all components in $r_\mathrm{al}[n]$ are aligned, the dispersed signal $\tilde{x}[n]$ can be efficiently removed through sample-wise division in the time domain. This results in a compensated signal, expressed as:
\begin{align}
    r_\mathrm{comp}[n]=\frac{\tilde{r}_\mathrm{al}[n]}{\tilde{x}[n]}
    =\sum_{l=0}^{L-1}a_\ell\e^{\j\varphi_\ell} \e^{\j2\pi f_\mathrm{b}nT_\mathrm{s}}+\frac{\breve{w}[n]}{\tilde{x}[n]}\text{.} \label{eq:s_comp}
\end{align}
Subsequently, the beat frequencies $f_{\mathrm{b,}\ell}$ can be estimated as outlined in (\ref{eq:RDM}), similar to traditional radar systems.

%% file: figures/spectrogram.tex
\centering
\begin{tikzpicture}
    \begin{axis}[width=0.8\columnwidth,
      name=main1,
      ylabel={Normalized power (dB)},
      xlabel={$f/f_\mathrm{s}$}, 
      xlabel style={sloped},
      ylabel style={sloped},
      legend style={
        legend cell align=left,
        legend columns=1,
        mark size=.6pt,
        font=\small,
        xmin=-0.5,
        xmax=0.5,
    },
      xmin=-0.5,
      xmax=0.5,
      xtick={-0.5,0,0.5}
      ]

      \addplot [KITcyanblue, thick]
      table[x=freq,y=X, col sep=comma] {figures/spectrums.csv};
      \addlegendentry{Radar};
      \addplot [KITpurple,  thick]
      table[x=freq,y=Xmod, col sep=comma] {figures/spectrums.csv};
      \addlegendentry{ISAC};
      \addplot [KITgreen, thick]
      table[x=freq,y=X_rvpc, col sep=comma] {figures/spectrums.csv};
      \addlegendentry{RVPC};
    \end{axis}

  \end{tikzpicture}

%% file: dispersion_sketch.tex
    \centering
    \begin{tikzpicture}[auto, line width=1pt, x=3.5em,y=2.9em]
        \draw[black, arrows = {-Latex[width = 1.2mm, length=1.2mm]}, thick] (0,0) to (0,4) node[left]{$f_\mathrm{beat}$};        
        \draw[black, arrows = {-Latex[width = 1.2mm, length=1.2mm]}, thick] (-0.4,0) to (4,0) node[below]{$\tau$};        
        \draw[kit-gray70 , thick] (0,0) to (3.5,3.5) node[below right]{$\tau_\mathrm{gr}=-\frac{f_\mathrm{b}}{\alpha}$};
        \draw[KITgreen, fill=kit-green70](0.2,2.2)--(1.4, 2.2)--(1.0, 2.6) node[right, KITgreen]{$r_\mathrm{al}(t)$} -- (-0.2,2.6) --cycle;
        \draw[KITgreen, fill=kit-green70](0.2,1.4)--(1.4, 1.4)--(1.0, 1.8)  -- (-0.2,1.8) --cycle;
        \foreach \y in {0.4,0.8,...,3.2}
            {\draw[kit-gray70, arrows = {-Latex[width = 1.2mm, length=1.2mm]}, thick] (\y, \y) to (0,\y);}

        \draw[kit-gray70] (1.5,3.3) node{$G_\mathrm{al}(f)=\e^{\j\frac{\pi f ^2}{\alpha}}$};
        \draw[KITcyanblue, fill=kit-cyanblue70](1.6,1.4) rectangle (3, 1.8) node[right, KITcyanblue]{$r_\mathrm{IF}(t-\tau)$}; 
        \draw[KITcyanblue, fill=kit-cyanblue70](2.4,2.2) rectangle (3.6, 2.6);

    \end{tikzpicture}

%% file: 3Performance_Analysis.tex
\renewcommand{\thefigure}{6}
\begin{figure}
    \centering
    \input{figures/histogram.tex} 
    \caption{Sent signal before and after dispersion through the alignment filter $G_\mathrm{al}([k])$ for QPSK and 16-QAM respectively.}
    \label{fig:Dispersed Signals}   
\end{figure}
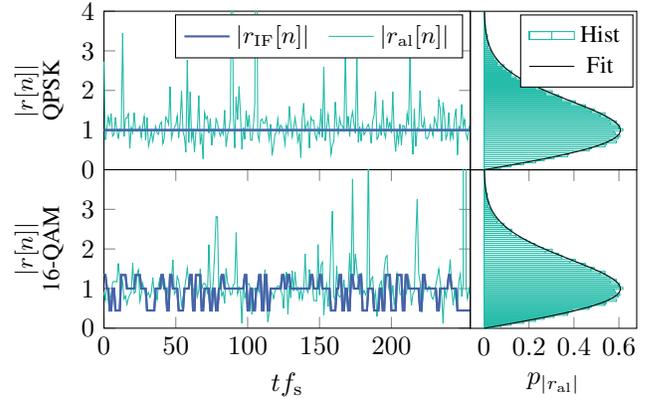

As can be inferred from (\ref{eq:s_comp}), the compensation of the information signal may enhance the noise in $r_\mathrm{comp}[n]$ depending on the stochastic characteristics of $\tilde{x}[n]$. In Fig. \ref{fig:spectrogram} we observe that the noise floor of the signal ater the compensation using the RVPC algorithm (green) is higher than the noise floor of the radar-only signal (blue). The performance degradation caused by this phenomenon was observed in \cite{Lampel_2022}, but not analyzed quantitatively. In this section, we derive the impact of the noise enhancement on the overall system performance.

\subsection{Sensing performance}
One important performance metric for sensing systems is the \ac{ISNR}, since it directly correlates with the ability of the system to detect and evaluate targets precisely through detection algorithms such as the \ac{CFAR} algorithm \cite{richards2005}. Fore each reflection $\ell$, the \ac{ISNR} is defined as 
\begin{align*}
    \mathrm{ISNR}_\ell=\frac{NP_{\mathrm{Peak,}\ell}}{P_\mathrm{N}}\text{,}
\end{align*}
where $P_{\mathrm{Peak,}\ell}=|a_\ell|^2$ corresponds to the peak power associated to a reflection $\ell$ and $P_\mathrm{N}$ is the power of the noise in the estimated signal. Notice that $\mathrm{ISNR}_\ell$ may vary for different targets within the same received signal depending on their individual amplitude $a_\ell$. In the following, we evaluate for both effects. For simplicity, we consider a single target scenario ($L = 1$) and drop the index $\ell$ in our derivation. This assumption does not compromise generality, as all operations involved are linear and can thus be applied to an arbitrary number $L$ of echos. 

Two main effects dominate the ISNR behavior of the complete system:
\begin{itemize}
    \item Firstly, the FFT leads to a coherent integration of the target power in (\ref{eq:s_comp}) into a single frequency bin, while the noise is added incoherently among all bins. This leads to a processing gain equal to the length $N$ of the FFT. 
    \item The noise $\frac{w[n]}{\tilde{x}[n]}$ in (\ref{eq:s_comp}) is enhanced by the division. In order to quantify this effect, we present a more detailed look into the spectrum and stochastic characteristics of $\tilde{x}[n]$.
\end{itemize}

Therfore, we consider the FFT of $r_\mathrm{IF}[n]$ in (\ref{eq:G_implementation}):
\begin{align}
    R_\mathrm{IF}[k]&=\mathrm{FFT}\mleft\{ r_\mathrm{IF}[n]\mright\}\\  %
                    &=aX[k-T_s f_{\mathrm{beat}}]\cdot \e^{\j2\pi \frac{\tau}{T_\mathrm{S}}k}+\check{W}[k]\text{.}
\end{align}
The spectrum of $X[k]$ can be decomposed into the pulse shape and the random symbols $\RV{c}_m$, leading to 
\begin{align*}
    R_\mathrm{IF}[k]=a\cdot G_\mathrm{ps}\mleft( k - T_sf_{\mathrm{beat}}\mright)\underbrace{\sum_{m=0}^{M-1}\RV{c_m}\e^{\j2\pi\frac{T_\mathrm{ch}}{M}mk}}_{\RV{Q}_k \sim \mathcal{CN} \mleft(0, \sigma^2_\RV{Q}\mright)}+ \check{W}[k]\text{,}
\end{align*}
where $G_\mathrm{ps}[k]$ is the frequency domain representation of the pulse shape filter $g_\mathrm{ps}[n]$.

Note that the FFT effectively performs the sum of $M$ independent random variables $\RV{c}_m$ weighted by a complex factor $\e^{\j2\pi\frac{T_\mathrm{ch}}{M}mk}$. According to the central limit theorem, the probability density function of this sum converges to a complex normal distribution $\RV{Q}_k\sim\mathcal{CN}\mleft(0, \sigma^2_\RV{Q}\mright)$ \cite{Billingsley86,Geiger25JCS}, where $\sigma^2_\RV{Q}$ corresponds to the variance of the added random variables, which in this case is given by $\mathrm{var}\{\RV{c}\}=\mathbb{E}\{\left| \RV{c}\right| ^2 \}=1$. Note that the actual constellation $\mathcal{C}$ does not have any statistically significant effect on the characteristics of this distribution. 

The alignment filter $G_\mathrm{al}[k]$ introduces an additional phase shift to each frequency component, which causes dispersion, but does not alter the stochastic characteristics of $\RV{Q}_k$.  This way, applying the IFFT in (\ref{eq:G_implementation}) yields
\begin{align*}
    r_\mathrm{al}[n]=\underbrace{\sum_{k=0}^{N-1} G_\mathrm{ps}\mleft( k - T_s f_{\mathrm{beat,}\ell}\mright)\cdot \RV{Q}_k \cdot \e^{\j\pi \frac{k^2}{\alpha}}\cdot \e^{\j \frac{2\pi}{N} nk}}_{\RV{q}_n\sim \mathcal{CN} \mleft(0, \sigma^2_\RV{q}\mright)}+ \breve{W}[k]  \text{.}
\end{align*}

 The behavior of the output random variable $\RV{q}_n$ heavily depends on the effect of the alignment filter $G_\mathrm{al}[k]$. We take a closer look at two extreme cases:

\subsubsection{Large chirp slope $\alpha$} If the chirp slope $\alpha$ is high compared to the bandwidth of $X[k]$, the alignment filter converges to $G_\mathrm{al}[k]\to 1$ yielding a negligible dispersive effect on the communication signal and resulting in 
\begin{align*}
    \lim_{\alpha \to \infty}\tilde{x}[n] = \mathrm{IFFT}\mleft\{\mathrm{FFT}\mleft\{x[n]\mright\}G_\mathrm{al}[k]\mright\} \stackrel{G_\mathrm{al}[k]\to 1}{=} x[n]
\end{align*}
In this case, the statistical characteristics of $x[n]$ dominate the noise enhancement in (\ref{eq:s_comp}). Previous literature, e.g., \mbox{\cite{Lampel_2022,Uysal_2020}} generally focuses on this case.  For signals with constant magnitude like PC-FMCW,  $|\tilde{x}[n]| \approx |x[n]|=1$ holds and the noise enhancement can be neglected.

\subsubsection{Small chirp slope $\alpha$}
For smaller slopes $\alpha$, the IFFT operation has to be analyzed as the sum of $N$  random variables $\RV{Q}_k$ with different complex weights. Once again, the components generated by the IFFT correspond to a random variable $\RV{q}_n \sim \mathcal{CN}\mleft(0, \sigma^2_\RV{Q}\mright)$ for each sample, following a normal distribution with the same mean and variance as $\RV{Q}_k$. All in all, only the power of the input constellation $\mathcal{C}$, which is normalized to $1$, has an impact on the noise enhancement, which neither depends on the actual symbols nor their individual magnitude. The magnitude of the transmit signal $x[n]$ in the time domain for a constant envelope signal (QPSK) and a higher order constellation (16-QAM) are shown in Fig.~\ref{fig:Dispersed Signals} together with the resulting magnitude of $\tilde{x}[n]$ after the alignment signal and the  histogram of the magnitudes in both cases. As expected, both follow a Rayleigh distribution with $\sigma_\mathrm{Ray}^2=1$. 

The noise term $\tilde{w}[n]=\frac{\breve{w}[n]}{\tilde{x}[n]}$ corresponds to the quotient of two normal distributed random complex variables, which results in a Cauchy-Lorentz distribution \cite{Baxley2010}. The variance of this distribution cannot be expressed analytically.
\subsubsection{General case}
For every case between the two extremes analyzed above, the distribution of $\tilde{x}[n]$ results in a mixture between the distribution of the input signal $x[n]$ and a complex normal distribution $\mathcal{CN}(0, \mathrm{var}(\mathcal{C}))$, rendering the derivation of analytical expression for the enhanced noise $\tilde{w}[n]$ infeasible. For this reason, we present a numerical evaluation of the overall ISNR for a wide set of parameters.

%% file: figures/histogram.tex
\begin{tikzpicture}[
    /pgfplots/scale only axis,
    /pgfplots/width=6cm,
    /pgfplots/height=6cm
]

\begin{axis}[
    name=main axis, %
    xmin=0,xmax=255,
    ymin=0,
    ymax=4,
    ytick={0, 1, 2,3, 4},
    width=0.55\columnwidth,
    height=6em,
    xtick=\empty,
    ylabel style={align=center, font=\small},
    ylabel={$\left| r[n] \right|$\\ $\text{QPSK}$},
    legend style={
        anchor=north east,
        legend columns=-1,
        /tikz/every even column/.append style={column sep=0.5em},
        font=\small
    }
]

\addplot [thick, KITblue]table [x=t, y=stream, col sep=comma, mark=none,]  {figures/stream_dispersed_bpsk.csv};
\addlegendentry{$\left|r_\mathrm{IF}[n]\right|$}
\addplot [kit-green70,]table [x=t, y=rvpc, col sep=comma, mark=none,]  {figures/stream_dispersed_bpsk.csv};
\addlegendentry{$\left|r_\mathrm{al}[n]\right|$}
\addplot [thick, KITblue]table [x=t, y=stream, col sep=comma, mark=none,]  {figures/stream_dispersed_bpsk.csv};
\end{axis}

\begin{axis}[
    anchor=north west,
    at=(main axis.north east),
    ytick=\empty,
    xtick=\empty,
    ymin=0,
    ymax=4,
    width=0.25\columnwidth, 
    height=6em,
    legend style={
        anchor=north east,
        legend columns=1,
        mark size=.6pt,
        font=\small
    }
]
\addplot [  xbar,
            bar width=0.05,
            kit-green70,
] table [y=abs, x=counts, col sep=comma, mark=none,]  {figures/hist_dispersed.csv};
\addlegendentry{Hist}
\addplot [ 
] table [y=abs, x=rayleigh, col sep=comma, mark=none,]  {figures/hist_dispersed.csv};
\addlegendentry{Fit}
\end{axis}

\begin{axis}[
    anchor=north west,
    at=(main axis.south west),
    xmin=0,xmax=255,
    ymin=0,
    ymax=4,
    width=0.55\columnwidth,
    height=6em,
    xlabel=$t f_\mathrm{s}$,
    xtick={0,50,...,200},
    ytick={0, 1, 2,3},
    ylabel style={align=center, font=\small},
    ylabel={$\left| r[n] \right|$\\ $\text{16-QAM}$}
]

\addplot [thick, KITblue]table [x=t, y=stream, col sep=comma, mark=none,]  {figures/stream_dispersed.csv};

\addplot [kit-green70,]table [x=t, y=rvpc, col sep=comma, mark=none,]  {figures/stream_dispersed.csv};
\addplot [thick, KITblue]table [x=t, y=stream, col sep=comma, mark=none,]  {figures/stream_dispersed.csv};
\end{axis}

\begin{axis}[
    anchor=north west,
    at=(main axis.south east),
    ytick=\empty,
    ymin=0,
    ymax=4,
    width=0.25\columnwidth, 
    height=6em,
    xlabel=$p_{|r_\mathrm{al}|}$
]
\addplot [  xbar,
            bar width=0.05,
            kit-green70,
] table [y=abs, x=counts, col sep=comma, mark=none,]  {figures/hist_dispersed.csv};

\addplot [ 
] table [y=abs, x=rayleigh, col sep=comma, mark=none,]  {figures/hist_dispersed.csv};

\end{axis}

\end{tikzpicture}

%% file: 4results.tex
To evaluate the overall sensing performance, we simulate a system with typical parameters of modern FMCW radar systems operating in the W-band. The carrier frequency is set to $\SI{77}{\giga\hertz}$. The chirp sweep bandwidth is $B_\mathrm{ch} = \SI{2}{\giga\hertz}$, and the sampling frequency is $f_\mathrm{s} = \SI{200}{\mega\hertz}$. We carry out a sweep of the  chirp slope $\alpha$ and corresponding chirp duration $T_\mathrm{ch}$.  In order quantify and compare the performance of different parameter sets, we introduce the normalized chirp slope $\alpha_\mathrm{norm}$, which relates the physical slope $\alpha$ to the sampling rate $f_\mathrm{s}$ and the amount of samples $N$ within a chirp period $T_\mathrm{ch}$. 
\vspace{-0.3cm}
\begin{align}
    \alpha_\mathrm{norm}=\frac{\alpha}{f_\mathrm{s}^2}=\frac{B/f_\mathrm{s}}{N}\text{.}
\end{align}
Fig. \ref{fig:kl_div}  shows the Kullback-Leibler-Divergence $D_\mathrm{KL}$ between the distribution of the dispersed samples $\tilde{x}[n]$ and the complex normal distribution of the random variable $\RV{q}_n\sim\mathcal{CN}(0, \mathrm{var}(\mathcal{C}))$ that results for small slopes, depending on $\alpha_\mathrm{norm}$. Additionally, we vary the effective symbol rate $R$ between \SI{50}{\mega Bd} and \SI{12.5}{\mega Bd} by upsampling and interpolating through a root raised cosine filter with roll-off factor $\beta=0.3$. 
As expected, the symbol rate $R$ and chirp slope $\alpha_\mathrm{norm}$ have the most significant effect on the distribution of the dispersed samples $\tilde{x}[n]$. We observe that $D_\mathrm{KL}(p_{\tilde{x}}\| \mathcal{CN}(0, \mathrm{var}(\RV{c}_m)))$ takes very small values for high symbol rates indicating a significant dispersive effect caused by $G_\mathrm{al}[k]$. In contrast, the modulated constellation $\mathcal{C}$ has a smaller effect on this behavior. 
\renewcommand{\thefigure}{7}
\begin{figure}
    \centering
\input{figures/kl_div.tex}
\vspace{-0.7cm}
\caption{Kullback-Leibler-Divergence $D_\mathrm{KL}(p_{\tilde{x}}\| \mathcal{CN}(0, \mathrm{var}(\RV{c}_m)))$ depending on the normalized chirp slope $\alpha_\mathrm{norm}$ and the symbol rate.}
\label{fig:kl_div}
\end{figure}
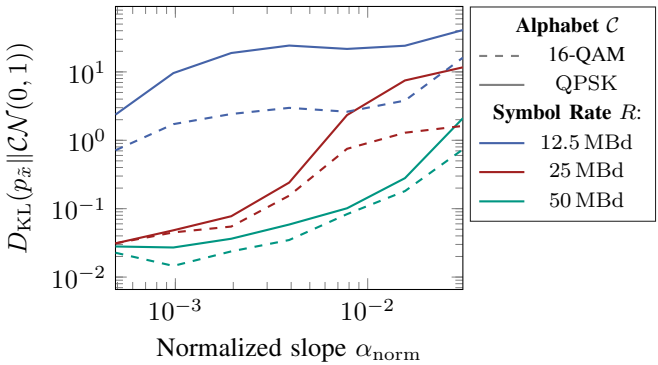

Additionally, we evaluate the overall sensing performance by calculating the ISNR for a parameter set with chirp duration $T_\mathrm{ch} = \SI{51,2}{\micro\second}$ i.e. $N=1024$ samples and a symbol rate of \SI{50}{\mega Bd}. This parameters correspond to a normalized chirp slope of $\alpha_\mathrm{norm}=10^{-3}$. The modulation order $\mleft|\mathcal{C}\mright|$ is additionally swept between 4 (QPSK) and 256-QAM, using square-shaped equidistant symbols. The SNR of the received signal is varied from $\SI{-20}{\decibel}$ to $\SI{+20}{\decibel}$. As a baseline, we compare the performance to that of a traditional FMCW radar waveform without communication capabilities. The resulting ISNR for both systems is shown Fig. \ref{fig:ISNR_Plot}.

The numerical simulations underline the behavior derived in Section \ref{sec:Performance}: Although there is a clear noise penalty enhancement of approximately \SI{10}{\decibel} due to the division by the dispersed signal $\tilde{x}[n]$, it is small compared to the processing gain achieved through the calculation of the periodogram.
Additionally, we notice that the order of the constellation has no significant impact on the sensing performance of the system. The phase coded case, corresponding to the constellation order 4 in Fig. \ref{fig:ISNR_Plot} does not lead to a better sensing performance than higher QAM modulation schemes. This enables the use of magnitude as a degree of freedom for communication, leading to a more efficient use of the physical resources.%

\renewcommand{\thefigure}{8}
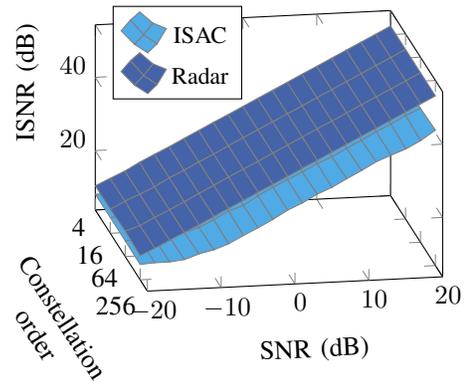
\begin{figure}
    \centering
\input{figures/isnr_3d}
\caption{ISNR of the sensing receiver for different modulation orders, for a chirp duration of $N=1024$ samples, $\alpha_\mathrm{norm}=10^{-3}$ and $R=\SI{50}{\mega Bd}$. }
\label{fig:ISNR_Plot}
\end{figure}

%% file: figures/kl_div.tex
\begin{tikzpicture}
    \begin{axis}[width=0.7\columnwidth,
      name=main,
      view={80}{24},
      zlabel={ISNR (dB)},
      ylabel={$D_\mathrm{KL}(p_{\tilde{x}}\| \mathcal{CN}(0,1))$},
      xlabel={Normalized slope $\alpha_\mathrm{norm}$}, 
      xmode=log,
      ymode=log,
      xmin=1/2048,
      xmax=1/32,
      legend style={ legend cell align=center, align=center, draw=white!15!black, font=\footnotesize, anchor=north west,
      at={(1.02,1)}
    },]
   
    \addlegendimage{white} \addlegendentry{
      \hspace{-6mm} \textbf{Alphabet} $\mathcal{C}$}
    \addlegendimage{gray, dashed, thick} \addlegendentry{16-QAM}

    \addlegendimage{gray, solid, thick} \addlegendentry{$\mathrm{QPSK}$}
    \addlegendimage{white} \addlegendentry{\hspace{-6mm} \textbf{Symbol Rate $R$}:}
    \addlegendimage{KITblue, thick} \addlegendentry{\SI{12.5}{\mega Bd}}
    \addlegendimage{KITred, thick} \addlegendentry{\SI{25}{\mega Bd}}
    \addlegendimage{KITgreen, thick} \addlegendentry{\SI{50}{\mega Bd}}

      \addplot [KITgreen, thick]
      table[y=qpsk4, x=slope, col sep=comma] {figures/kl_div_rrc.csv};
      \addplot [KITgreen, dashed, thick]
      table[y=qam4, x=slope, col sep=comma] {figures/kl_div_rrc.csv};
      \addplot [KITred, thick]
      table[y=qpsk8, x=slope, col sep=comma] {figures/kl_div_rrc.csv};
      \addplot [KITred, dashed, thick]
      table[y=qam8, x=slope, col sep=comma] {figures/kl_div_rrc.csv};
      \addplot [KITblue, thick]
      table[y=qpsk32, x=slope, col sep=comma] {figures/kl_div_rrc.csv};
      \addplot [KITblue, dashed, thick]
      table[y=qam32, x=slope, col sep=comma] {figures/kl_div_rrc.csv};

    \end{axis}

  \end{tikzpicture}

%% file: figures/isnr_3d.tex
\begin{tikzpicture}
    \begin{axis}[width=0.7\columnwidth,
      height=0.6\columnwidth,
      name=main,
      view={80}{24},
      zlabel={ISNR (dB)},
      xlabel style={sloped, align=center},
      ylabel style={sloped},
      xlabel={Constellation \\ order},
      ylabel={SNR (dB)}, 
      legend style={
         at={(0.05,0.85)},
        anchor=west,
        legend columns=1,
        mark size=.6pt,
        font=\small,
    },
    xmax=500,
    xmode=log,
    xtick={4,16,64,256},
    xticklabels={4, 16 , 64, 256}
      ]
      \addlegendentry{ISAC}
      \addplot3 [surf, mesh/rows=4,faceted color=gray,
      fill=KITcyanblue, fill opacity=0.9]
      table[y=snr, x=M_Sym, z=ISNR, col sep=comma] {figures/stats512.csv};
   
      \addplot3 [surf, mesh/rows=4, faceted color=gray, 
      fill=KITblue, fill opacity=0.8]
      table[y=snr, x=M_Sym, z=ISNR_unmod, col sep=comma] {figures/stats512.csv};
      \addlegendentry{Radar}
    \end{axis}

  \end{tikzpicture}

%% file: 5conclusions.tex
In this paper, we provide a comprehensive overview of the key signal processing steps in a FMCW-ISAC system, detailing their implementation as digital discrete systems. We analyze the impact of every signal processing step on the sensing performance and present a quantitative study of the stochastic properties of relevant signals at each stage.

We additionally propose and evaluate higher order modulations with non-constant magnitude in FMCW-ISAC systems. Our findings reveal that the order of the constellation does not have a substantial impact on the ISNR as compared to the sensing signal processing gain. This insight highlights the potential advantages of using higher-order constellations in automotive radar systems, as they enable larger information throughput in ISAC systems without degrading sensing performance.